\documentclass[aps,prl,twocolumn,floatfix,showpacs]{revtex4}
\usepackage{epsf,amsmath,amssymb,wasysym}
\newcommand{\be}{\begin{equation}}
\newcommand{\ee}{\end{equation}}

\begin{document}
\title{Burstiness and Memory in Complex Systems}
\author{Kwang-Il Goh$^{1,2}$ and Albert-L\'aszl\'o Barab\'asi$^1$} 
\affiliation{$^1$Center for Complex Network Research and Department of Physics, 225 Nieuwland Science Hall, University of Notre Dame, Notre Dame, IN 46556, USA
\\
$^2$Department of Physics, Korea University, Seoul 136-713, Korea
}
\date{\today}
\begin{abstract}
The dynamics of a wide range of real systems, from email patterns to
earthquakes, display a bursty, intermittent nature, characterized by
short timeframes of intensive activity followed by long times of no or
reduced activity. The understanding of the origin of such
bursty patterns is hindered by the lack of tools to compare different
systems using a common framework. We introduce two measures
to distinguish the mechanisms responsible for the 
bursty nature of real signals, changes in the interevent times
and memory. 
We find that while the burstiness of natural phenomena
is rooted in both the interevent time distribution 
and memory, for human dynamics memory is weak, 
and the bursty character is due to changes in
the interevent time distribution.
Finally, we show that current models lack in their ability 
to reproduce the activity pattern observed in real systems,
opening up new avenues for future work.
\end{abstract}
\pacs{89.75.-k, 05.45.Tp}
\maketitle

The dynamics of most complex systems is driven by the loosely 
coordinated activity of a large number of components, such as 
individuals in the society or molecules in the cell.
While we witnessed much progress in the study of 
the networks behind these systems \cite{reviews},
advances towards understanding the system's dynamics has been 
slower.  With increasing potential 
to monitor the time-resolved activity of most components
of selected complex systems, such as
time-resolved email \cite{eckmann,alb,vazquez},
web browsing \cite{dezso}, and gene expression \cite{golding} patterns,
we have the opportunity to ask an important question: 
is the dynamics of complex systems governed by generic organizing 
principles, or each system has its own distinct dynamical features? 
While it is difficult to offer a definite answer to this question, 
a common feature across many systems is increasingly documented:
the burstiness of the system's activity patterns. 

Bursts, vaguely  corresponding to significantly enhanced activity levels 
over short periods of time followed by long periods of 
inactivity, have been observed in a wide range of systems, 
from email patterns \cite{alb} to earthquakes \cite{bak,corral} 
and gene expression \cite{golding}.
Yet, often burstiness is more of a metaphor than 
a quantitative feature, and opinions about its origin diverge.
In human dynamics, burstiness has been reduced to the fat-tailed 
nature of the response time distribution \cite{alb,vazquez}, in contrast with
earthquakes and weather patterns, where memory effects appear to 
play a key role \cite{bunde,livina}. 
Once present, burstiness can affect 
the spreading of viruses \cite{alb} or resource allocation \cite{willinger}. 
Also, deviations towards regular,
``anti-bursty'' behavior in heartbeat may indicate 
disease progression \cite{thurner}.  Given the diversity 
of systems in which it emerges, there is a need to place burstiness 
on a strong quantitative basis. Our goal in this paper 
is to make a first step in this direction,
by developing measures that can help quantify 
the magnitude and potential origin of the bursty patterns seen in
different real systems.

Let us consider a system whose components 
have a measurable activity pattern that can be mapped 
into a discrete signal, recording the moments when some events take place, 
like an email being sent, or a protein being translated. 
The activity pattern is random (Poisson process)
if the probability of an event is time-independent. 
In this case the interevent time between two consecutive events 
$(\tau)$ follows an exponential distribution,
$P_{\textrm{P}}(\tau)\sim\exp(-\tau/\tau_0)$ (Fig.~1a). 
An apparently bursty (or anti-bursty) signal 
could emerge if $P(\tau)$ is different from the exponential,
such as the bursty pattern of Fig.~1b,
or the more regular pattern of Fig.~1c.
Yet, changes in the interevent time distribution is not
the only way to generate a bursty signal.
For example, the signals shown in Fig.~1d,e have exactly
the same $P(\tau)$ as in Fig.~1a, 
yet they have a more bursty or a more regular character. 
This is achieved by introducing memory:
in Fig.~1d the short interevent times tend to follow short ones, 
resulting in a bursty look. 
In Fig.~1e the relative regularity is due to 
a memory effect acting in the opposite direction: 
short (long) interevent times tend to be followed by long (short) ones.
Therefore, the apparent burstiness of a signal 
can be rooted in two, mechanistically quite
different deviations from a Poisson process:
changes in the interevent time distribution or memory.
To distinguish these effects, we introduce
the burstiness parameter $\Delta$ and the memory parameter $\mu$, 
that quantify the relative contribution of each
in real systems. 

\begin{figure}[!ht]
\centerline{\epsfxsize=8.5cm \epsfbox{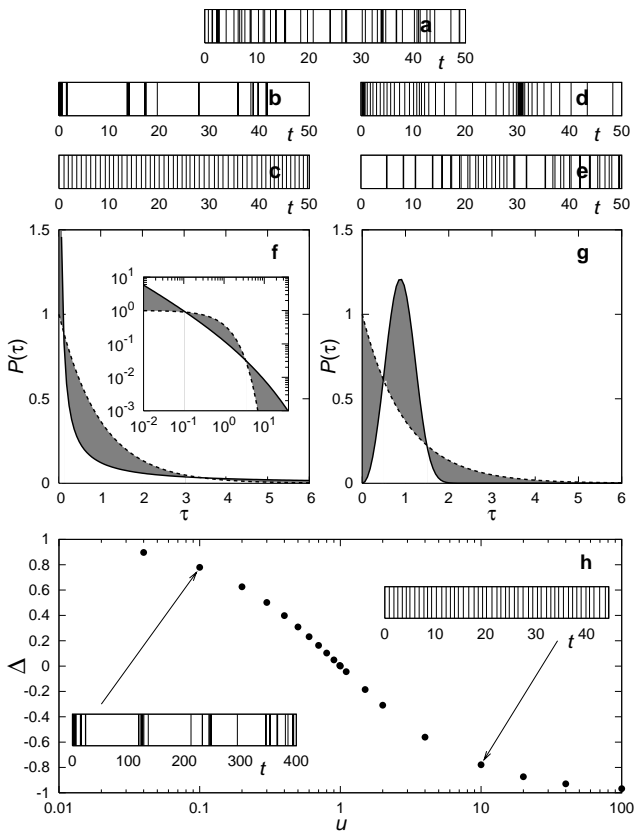}}
\caption{
(a) A signal generated by a Poisson process with a unit rate.
(b,c) Bursty character through the interevent time 
distribution: A bursty signal (b) generated by 
the power-law interevent time distribution $P(\tau)\sim\tau^{-1}$,
and an anti-bursty signal (c) generated by the Gaussian interevent
time distribution with $m=1$ and $\sigma=0.1$.
A bursty signal can emerge through memory as well.
For example, the bursty signal shown in (d) is obtained 
by shuffling the Poisson signal of (a) to increase the memory effect.
A more regular looking signal, with negative memory, is
obtained by the same shuffling procedure (e). Note that signals in (a), (d) and (e) have 
identical interevent time distribution.
(f) An interevent time distribution (solid line) will 
appear bursty $(\Delta>0)$ if it has a higher frequency of 
long or short interevent times than predicted for a Poisson
process (dotted line). Inset shows the same curves in
log-log scale.
(g) The signal will appear to be regular (anti-bursty, $\Delta<0$)
if $P(\tau)$ is higher in the average interevent time region
than a Poisson process.
The burstiness parameter $\Delta$ is 
half of the shaded area between the corresponding interevent time
distribution (solid) and the reference distribution (dotted).
(h) The stretched exponential interevent distribution 
interpolates between a highly bursty $(\Delta=1)$,
a Poisson $(\Delta=0)$, and a regular $(\Delta=-1)$ signal.
The figure shows $\Delta$ as a function of the parameter $u$.
}
\end{figure}

The {\em burstiness parameter} $\Delta$ is defined as
\be 
\Delta \equiv \frac{\textrm{sgn}(\sigma_{\tau}-m_{\tau})}{2}
\int_{0}^{\infty}|P(\tau) - P_{\textrm{P}}(\tau)|d\tau~, 
\ee
where $m_{\tau}$ and $\sigma_{\tau}$ are the mean and
the standard deviation of $P(\tau)$ \cite{alexei}. 
The meaning of $\Delta$ is illustrated in Fig.~1f--h, where we
compare $P(\tau)$ for a bursty- (Fig.~1f) and an anti-bursty (Fig.~1g) 
signal with a Poisson interevent time distribution.
A signal will appear bursty if the frequency of the short 
and long interevent times is higher than in a random signal (Fig.~1f),
resulting in many short intervals separated by longer periods of
inactivity. Therefore, there are fewer interevent times
of average length than in a Poisson process.
A signal displays anti-bursty character, however, 
if the frequency of the interevent times is 
enhanced near the average and depleted
in the short and long interevent time region (Fig.~1g).
$\Delta$ is bounded in the range $(-1,1)$, and its magnitude
correlates with the signal's burstiness: $\Delta=1$ is the most bursty signal,
$\Delta=0$ is completely neutral (Poisson), and $\Delta=-1$ corresponds
to a completely regular (periodic) signal. 
For example, in Fig.~1h we show $\Delta$
for the stretched exponential distribution, 
$P_{\textrm{SE}}(\tau) = u(\tau/\tau_0)^{u-1}\exp[-(\tau/\tau_0)^u]/\tau_0~,$
often used to approximate the interevent time distributions
of complex systems \cite{laherrere}.
The smaller the $u$ is, the burstier is the signal, and for
$u\to0$, $P_{\textrm{SE}}(\tau)$ follows 
a power law with the exponent $-1$, for which $\Delta=1$.
For $u=1$, $P_{\textrm{SE}}(\tau)$ is simply the exponential distribution 
with $\Delta=0$.
Finally, for $u>1$, the larger $u$ is, the more regular is the signal,
and for $u\to\infty$, $P(\tau)$ converges to a Dirac delta function with $\Delta=-1$.

Most complex systems display a remarkable heterogeneity: 
some components may be very active, and others much less so. 
For example, some users may send dozens of emails during a day, 
while others only one or two.  
To combine the activity levels of so different components,
we can group the signals
based on their average activity level, and measure $P(\tau)$ 
only for components with similar activity level.
As the insets in Fig.~2 show, the obtained curves are systematically
shifted. If we plot, however, $\tau_0 P(\tau)$ as a function of 
$\tau/\tau_0$, where $\tau_0$ being the average interevent time,
the data collapse into a single curve 
${\cal F}(x)$ (Fig.~2), 
indicating that the interevent time distribution follows 
$P(\tau)=(1/\tau_0){\cal F}(\tau/\tau_0)$, where ${\cal F}(x)$ 
is independent of the average activity level of the component, 
and represents an universal characteristic 
of the particular system \cite{corral,saichev}. 
This raises an important question: 
will $\Delta$ depend on $\tau_0$? 
The burstiness parameter $\Delta$ is invariant under the time rescaling
as $\tilde{\tau}\equiv\tau/\tau_0$ and $\tilde{P}(\tilde{\tau})\equiv\tau_0P(\tau)$ 
with a constant $\tau_0$, since 
$\tilde{\Delta}\equiv\int_0^{\infty}|\tilde{P}(\tilde{\tau})-\tilde{P}_0(\tilde{\tau})|d\tilde{\tau}
=\int_0^{\infty}|\tau_0P(\tau)-\tau_0P_0(\tau)|d(\tau/\tau_0)
=\int_0^{\infty}|P(\tau)-P_0(\tau)|d\tau\equiv\Delta$,
{\it i.e.}, it characterizes the universal function ${\cal F}(x)$.
Such invariance of $\Delta$ enables us to 
assign to each system a single burstiness parameter, despite the 
different activity level of its components.

\begin{figure*}
\centerline{\epsfxsize=\linewidth \epsfbox{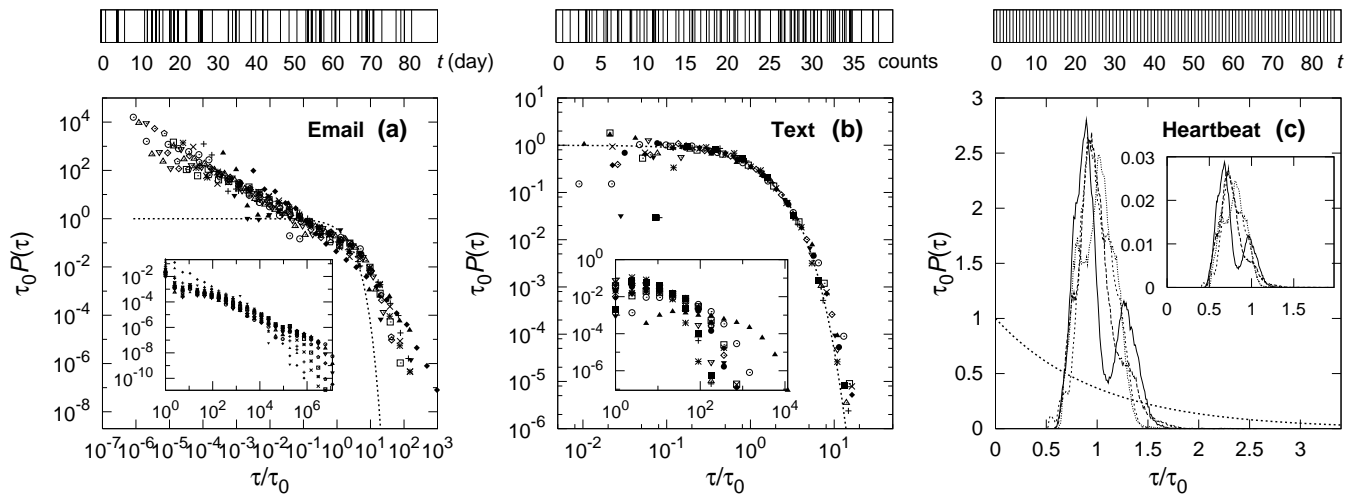}}
\caption{
Interevent time distributions $P(\tau)$ for some real signals.
(a) $P(\tau)$ for e-mail activity of individuals 
from a University \cite{eckmann}. $\tau$ corresponds
to the time interval between two emails sent by the same user.
(b) Interevent time distribution for the occurrence of letter in the text of 
C. Dickens' {\em David Copperfield} \cite{gutenberg}. 
(c) Interevent time distribution of cardiac rhythm of individuals \cite{physiobank}.
Each event corresponds to the beat in the heartbeat signal.
In each panel, we also show for reference the exponential interevent time
distribution (dotted). 
Unscaled interevent time distribution is shown in the inset for each
dataset.
}
\end{figure*}

The {\em memory coefficient} $\mu$ of a signal is defined as the
correlation coefficient of all consecutive interevent time values
in the signal over a population. 
That is, given all pairs of consecutive
interevent times $(\tau_{k,i},\tau_{k,i+1})$ for all components
$\{k=1,\cdots,N\}$,
\be
\mu \equiv \frac{1}{N}\sum_{k=1}^{N}\sum_{i=1}^{n_k-1}
\frac{(\tau_i-m_{{k1}})(\tau_{i+1}-m_{{k2}})}{\sigma_{{k1}}\sigma_{{k2}}}~,
\ee
where $N$ is the number of components in the system, 
$n_k$ is the number of events recorded for component
$k$, and $m_{k1} (m_{k2})$ and $\sigma_{k1} (\sigma_{k2})$
are the mean and standard deviation of $\tau_{k,i}$'s ($\tau_{k,i+1}$'s),
respectively.
The memory coefficient is positive when a short (long) interevent
time tends to be followed by a short (long) one,
and it is negative when a short (long) interevent time is likely
to be followed by a long (short) one.
The measurements indicate that $\mu$ is independent of $\tau_0$.

{\em Mapping complex systems on the $(\mu,\Delta)$ space---}
Given that the burstiness of a signal can have two
qualitatively different origins, the best
way to characterize a real system is to identify its
$\mu$ and $\Delta$ parameters, placing them
in a $(\mu,\Delta)$ phase diagram (Fig.~3).
As a first example, we measured the spacing between the consecutive 
occurrence of the same letter in texts of different
kind, era, and language \cite{gutenberg}. For these signals,
we find $\Delta\approx0$, {\it i.e.}, the interevent time distribution 
follows closely an exponential (Fig.~2b)
and $\mu\approx0.01$, indicating the lack of memory. 
Thus this signal is best described by a Poisson process,
at the origin of the phase diagram (Fig.~3).
In contrast, natural phenomena, like earthquakes \cite{japan} and weather
patterns \cite{rain} are in the vicinity of the diagonal,
indicating that $P(\tau)$ and memory equally
contribute to their bursty character.
The situation is quite different, however, 
for human activity, ranging from email and phone
communication to web browsing and library visitation patterns \cite{eckmann,
dezso,vazquez,callcenter}.
For these we find a high $\Delta$ and small or negligible $\mu$,
indicating that while these systems display significant
burstiness rooted in $P(\tau)$, 
memory plays a small role in their temporal inhomogeneity.
This lack of memory is quite unexpected, as it suggests
the lack of predictability in these systems in contrast with
natural phenomena, where strong memory effects could lead to
predictive tools.
Finally for cardiac rhythms describing the time between
two consecutive heartbeats (Fig.~2c) \cite{physiobank},
we find $\Delta_{\textrm{cardiac, healthy}}=-0.73(4)$ for 
healthy individuals and $\Delta_{\textrm{cardiac, CHF}}=-0.82(6)$
for patients with congestive heart failure (CHF),
both signals being highly regular.
Thus the $\Delta$ parameter captures the fact that 
cardiac rhythm is more regular with CHF
than in the healthy condition \cite{thurner}.
Furthermore, we find $\mu\approx 0.97$, 
indicating that memory also plays a very important role in the signal's
regularity.

The discriminative nature of the $(\mu,\Delta)$ phase diagram
is illustrated by the non-random placement of the different
systems in the plane: human activity patterns
cluster together in the high $\Delta$, low $\mu$ region, natural
phenomena near the diagonal, heartbeats in the high $\mu$, negative
$\Delta$ region and texts near $\Delta=\mu=0$, 
underlying the existence of distinct classes of 
dynamical mechanisms driving the temporal activity in these systems. 

\begin{figure*}[t]
\centerline{\epsfxsize=14cm \epsfbox{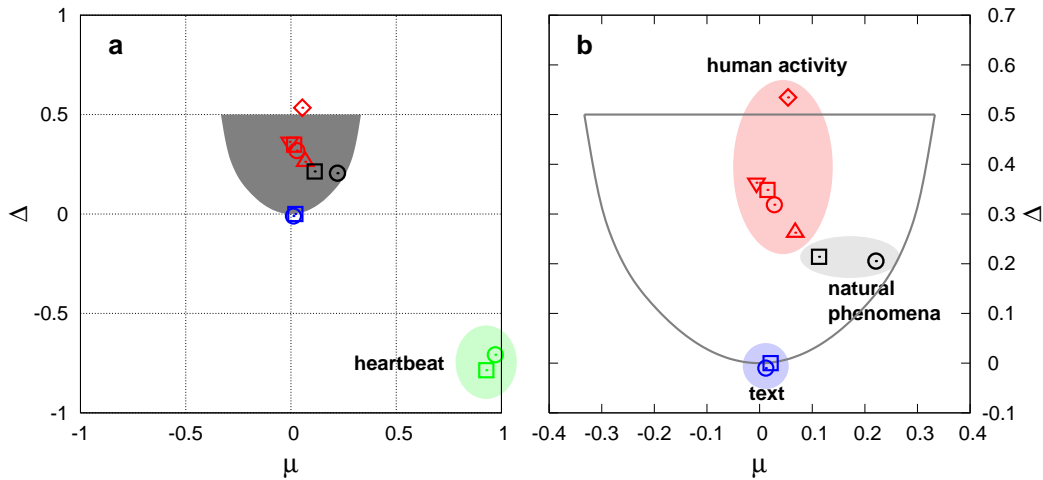}}
\caption{(Color) (a) The $(\mu,\Delta)$ phase diagram.
Human behaviors (red) are captured by activity patterns pertaining
to email ($\Box$) \cite{eckmann}, library loans ($\circ$) \cite{vazquez}, 
and printing ($\diamond$) \cite{paczuski} 
of individuals in Universities, 
call center record at an anonymous bank ($\triangle$) \cite{callcenter},
and phone initiation record from a mobile phone company ($\triangledown$).
Data for natural phenomena (black) are
earthquake records in Japan ($\circ$) \cite{japan} and
daily precipitation record in New Mexico, USA ($\Box$) \cite{rain}.
Data for human texts (blue) \cite{gutenberg} are
the English text of {\em David Copperfield} ($\circ$) and
the Hungarian text of {\em Isten Rabjai}
by G\'ardonyi G\'eza ($\Box$). 
Data for physiological behaviors (green) 
are the normal sinus rhythm ($\circ$) and 
the cardiac rhythm with CHF ($\Box$) of human subjects \cite{physiobank}.
Grey area is the region occupied by the 2-state model \cite{kleinberg}.
(b) Close-up of the most populated region.
}
\end{figure*}

Following the clustering of the empirical measurements in the
phase diagram, a natural question emerges: to what degree can current
models reproduce the observed quantitative features of bursty
processes? 
Queueing models, proposed to capture human activity patterns,
are designed to capture the waiting times of the tasks, 
rather than interevent times \cite{alb,vazquez,joao}.
Therefore, placing them on the phase diagram is not meaningful.
A bursty signal can be generated by 2-state model \cite{kleinberg},
switching with probability $p$
its internal state between Poisson processes with 
two different rates $\lambda_0 < \lambda_1$.
$\Delta$ for the model is independent of $p$ in the long time limit
as long as $p>0$, 
and takes its value in the range $0< \Delta<0.5$, approaching $0$ 
when $\lambda_0\approx\lambda_1$ 
and $0.5$ as $\lambda_1\to\infty$ and $\lambda_0\to0$.
The memory coefficient of the model 
follows $\mu=A(0.5-p)$, where $A$ is a positive
constant dependent on $\lambda_0$ and $\lambda_1$
so that $-1/3<\mu<1/3$.
The region in the $(\mu,\Delta)$ space occupied by the model
is shown in the light grey area in Fig.~3a,
suggesting that by changing its parameters the model could 
account for all observed behaviors.
Yet, the agreement is misleading: for example,
for human activities $P(\tau)$ is fat-tailed,
which is not the case for the model.
This indicates that $\Delta$ and $\mu$
offer only a first order approximation for
the origin of the burstiness, and for a detailed comparison
between models and real systems we need to inspect
other measures as well, such as the 
functional form of $P(\tau)$.
It also indicates the lack of proper modeling
tools to capture the detailed mechanisms 
responsible for the bursty interevent time 
distributions seen in real systems,
opening up possibilities for future work.

We would like to thank S. Havlin and A. V\'azquez
for helpful discussions.
This work is supported by the S.\ McDonnell Foundation and the
National Science Foundation under Grant No.\ CNS-0540348
and ITR DMR-0426737.


\begin{thebibliography}{99}
\bibitem{alexei} As an alternative, $\sigma_{\tau}/m_{\tau}$ can also
be used intead of $\Delta$ to meausre burstiness [A.\ V\'azquez, private
communications].
\end{thebibliography}
\end{document}